\documentclass[sigconf]{acmart}

\usepackage{booktabs} 

\setcopyright{none}

\acmDOI{}
\acmISBN{}
\acmConference[HaPoP 2018]{Fourth Symposium on the History and Philosophy of Programming}{March 2018}{Oxford, UK}
\copyrightyear{2018}

\begin{document}
\title{What can a 1980s BASIC programming textbook teach us today?}
\subtitle{Fourth Symposium on the History and Philosophy of Programming, March 2018, Oxford, UK}

\author{Martin Lester}
\affiliation{%
  \institution{University of Oxford}
}
\email{martin.lester@cs.ox.ac.uk}


\begin{abstract}

\emph{Elementary Basic}, published in 1982,
is an introductory programming text with a novel central conceit,
namely that the fictional 19th century detective Sherlock Holmes
used a computer to help solve mysteries.
It is also novel among similar books of its time for its focus on program design.
In other regards, such as its use of the language BASIC, it is representative of its time.

Over 35 years after it was written,
I think it is worth looking back at it to see to what is still relevant today
and what would be done differently.
We may even learn something about teaching programming today.
Of particular interest is the degree to which the use of BASIC
influenced the content of the book.

\end{abstract}

\maketitle

\section{Background}

The early 1980s were a revolutionary period in the development of home computing and computer programming.
For the first time, commercially produced home computers were available for sale to the general public at reasonable prices.
Most such computers came equipped with an interpreter for the programming language BASIC
that was loaded immediately from ROM when the computer was turned on.
The ease with which ordinary people could access a programming environment
helped to drive interest in learning programming and consequently the production of materials to support this:
newsagents routinely sold computer magazines that would include articles about programming in BASIC;
books giving tutorials on how to program were also available.

\section{Elementary Basic}

One such book was \emph{Elementary Basic: Learning to Program your Computer in Basic with Sherlock Holmes}~\cite{0394524233},
published in 1982.
The central conceit of the book is that Sherlock Holmes used Charles Babbage's Analytical Engine
to assist in solving various mysteries.
The authors have supposedly discovered unpublished manuscripts detailing
Watson's discussions with Holmes on these ventures into programming
and translated them into BASIC for the benefit of the modern reader.
(The cover of the book seems to claim that Babbage's Analytical Engine
came into use around the time the first Holmes story was written,
when in fact it was never built,
but given that Holmes is a fictional character,
the authors may be forgiven for suggesting he interacted with a non-existent machine.)

By the time the book was written and published,
the market for introductory programming books was already quite crowded.
In the postscript, the authors question
``whether the world really needed yet another text on programming'', but they believe that:
\begin{itemize}
\item ``programming is difficult\ldots [but] the basis of programming stems from a few elementary ideas'';
\item the dialogue format of the book is an ``easy to follow and enjoyable'' exposition of those ideas;
\item and ``the best method to teach programming is through problems''.
\end{itemize}
That is, the book was an attempt to teach introductory programming rather
than an attempt to teach a specific language.
This distinction may have been unclear to a potential buyer or reader and remains so today:
introductory books are necessarily usually tied to a single language,
which typically features prominently in the title.
As a more modern example, consider \emph{Objects First with Java}~\cite{9781292159041},
which similarly serves as a problem-based introductory programming text,
but does not distinguish itself from a language introduction through its title.

Most chapters of the book follow a set format:
\begin{itemize}
\item After some setup, Holmes discusses with Watson an aspect of a mystery that could be solved easily using a computer.
\item Holmes presents pseudo-code for solving the problem, followed by a BASIC implementation.
\item Watson asks questions about new ideas or confusing parts of the program, which Holmes answers.
\item The chapter concludes with an ``out of character'' summary of any new language features introduced.
\end{itemize}

\section{Observations}

I now outline some interesting observations about this book,
which I will expand upon in my talk.

\subsection{Concept}

The setting of the book, or perhaps just the fact that one can consider a programming text as having a setting,
is arguably its most distinctive feature.
Certainly, I found myself wanting to read it on this basis, despite having no particular interest in Sherlock Holmes
and little interest in BASIC in the last 15--20 years.
Surprisingly, as the book's acknowledgements reveal,
several people other than the credited authors played substantial roles in the development of the book's text,
in particular in writing the pastiches of Sherlock Holmes stories.

The use of dialogues as a means of exposition has a long history in Western science and philosophy,
for example in the works of Plato and Galileo,
although the style of the dialogues between Holmes and Watson is more like that between a master and a scholar.
An advantage of this format is that it allows errors and difficulties to be raised naturally by the character of the scholar;
a conventional narrative attempting to discuss points of misunderstanding risks seeming contrived or condescending.

\subsection{Learning by Solving Problems}

The idea of introducing programming through a series of concrete problems was certainly,
as the authors claimed, uncommon at the time.
Contemporary books were typically structured around the introduction of new language features,
with short programs illustrating their use~\cite{0856136662,0860206742}.

Although many of the problems in the book are motivated by the need to solve a mysterious murder or theft,
the actual programs largely end up being somewhat conventional.
Examples include calculating the date and month of the $n$th day of the year,
pretty-printing a coroner's report
and searching a flat file database of criminals for a matching record.

One exception is the first program in the book,
which purports to solve a logic puzzle of the kind one might find in a puzzle magazine,
with the aim of identifying a murderer.
I was intrigued by how a relatively complex problem such as this might be written and introduced in the book.
Upon looking at the program, the answer became clear.
Rather than being a general-purpose solver,
the program hardcoded all the data and inference rules, tying it completely to a specific problem instance.

When one considers all the difficulties that would be involved in writing a general-purpose solver in BASIC,
it is easy to see why this was a necessary simplification.
For example, the absence of any kind of pointers or references makes it difficult to construct any interesting data structure,
other than by allocating a large array and effectively managing the memory manually.

The verbosity of BASIC flow control also makes the programs much longer and harder to follow than would be necessary in other languages.
The majority of the example programs could be written in less than half the space in C or Java
and would perhaps take only a few lines in ML or Haskell.

Combining these factors, perhaps one can see why this and other books of its time
contain few algorithms of any significance.
For example, sorting may be a programming cliche, but I was surprised not to see it discussed at all
in the 250 pages of \emph{Elementary Basic}.
Overall, the mostly pedestrian examples and lack of algorithmic development
felt like a missed opportunity, given the book's unusual and promising concept.

\subsection{Program Development}

The credited authors of the book, Henry Ledgard and Andrew Singer,
are well-qualified as computer scientists,
and this shows through in their discussion of the design and correctness of programs.
Sadly, BASIC offers few facilities to support this and in some places
actively hinders it.

\subsubsection*{Types}
The idea of the distinction between types as enforced by the language and as
intended by the programmer is raised early in the book.
Types are given a more detailed discussion later on,
but given the paucity of BASIC's type system,
they are treated largely as something for the programmer to check when writing code,
rather than as something for the computer to check for the programmer.

\subsubsection*{Top-Down Design}
The book consistently advocates top-down design as a style of programming.
The choice of design technique is not as significant as the fact that it is discussed throughout and at great length,
which seems to be a rarity for an introductory book of the time.

\subsubsection*{Code Hygiene}
The text contains extensive reminders to use meaningful variable names where possible,
to add comments and to use indentation meaningfully.
Each of these prescriptions remains good and frequently ignored advice today.
Unfortunately, they could all be thwarted by certain BASIC implementations,
as the authors acknowledge.

\section{Conclusions}

\emph{Elementary Basic} had some novel ideas that would be worth revisiting today.
It would be interesting to see a modern introductory programming book (or perhaps a series of videos)
presented as a series of dialogues.
If done well, it could be very appealing to a young audience.
Computer scientists thinking about writing programming books for a broad audience should seriously
consider following this book's example by finding a skilled writer who can help to make the book interesting!

Learning to program by solving problems is pedagogically sound.
Looking at the complexity of problems in more recent books following this idea,
I argue we are getting better at writing programs, but this may be because we have better languages,
rather than because we are better at teaching how to use them,
although I hope our teaching has improved too.
(Note that we have better languages partly because we have more powerful computers that can support them.)

In the UK, in some of the more theoretically-oriented computer science degree programmes,
undergraduates are told that it does not particularly matter what language they learn to program in,
provided they learn the underlying principles.
While there is some truth in this claim,
I doubt it holds when the expressivity of the language becomes an impediment to writing interesting programs,
as I believe this book demonstrates is the case in BASIC.

Concerning the broader question of whether everyone in the future will be a programmer,
we sometimes look back nostalgically at the 1980s as a time of great computer literacy,
with large numbers of schoolchildren capable of programming to some degree.
Certainly, the ease with which one could access a BASIC interpreter and introductory programming materials was a major contributing factor to this.
But today, anyone who wishes can easily download (for example) a Python interpreter and textbook, or access an interactive online course.
Furthermore, looking at the technical content of this and other 1980s programming texts, I see little evidence of algorithmic sophistication.
I argue that most ``programmers'' of the 1980s were relatively superficial in their abilities
and most of what most of them achieved could be replicated today (perhaps using a database, spreadsheet or image editor) without ``programming''.
Contrary to most predictions, I therefore suggest that while the number of technically skilled people will increase in the future,
the number of programmers will decrease, or our understanding of what constitutes ``programming'' will change significantly.

%


\bibliographystyle{ACM-Reference-Format}
\bibliography{refs}

\end{document}